\documentclass[letter,11pt]{article}
\usepackage[top=1in,bottom=1in,left=1in,right=1in]{geometry}
\usepackage{slashed}
\usepackage{epsfig}
\usepackage{comment}
\usepackage{cancel}
\usepackage{bbm}
\usepackage{array}
\usepackage{bigints}
\usepackage{booktabs}
\usepackage{color}
\usepackage{dsfont}
\usepackage{float}
\usepackage{framed}
\usepackage{graphicx}
\usepackage{indentfirst}
\usepackage{mathrsfs}
\usepackage{multirow}
\usepackage{subdepth}
\usepackage{titlesec}
\usepackage[dotinlabels]{titletoc}
\usepackage{wrapfig}
\usepackage[all]{xy}
\usepackage[vcentermath]{youngtab}
\usepackage{relsize}
\usepackage{hyperref}
\numberwithin{equation}{section}

\usepackage[utf8]{inputenc}
\usepackage{slashed}
\usepackage{amsmath}
\usepackage{amsfonts}
\usepackage{amssymb}
\usepackage{cite}
\usepackage{mathtools}

\usepackage{cleveref}
\crefname{section}{§}{§§}
\Crefname{section}{§}{§§}

\def\ip{${\mathscr I}^+$}

\def\e{{\epsilon}}

 \def\p{\partial}
 \def\bz{{\bar z}}
 \def\bw{{\bar w}}
 
\def\0{{(0)}}
\def\1{{(1)}}
\def\2{{(2)}}

\def\<{\langle }
\def\>{\rangle }
\def\bw{{\bar w}}
\newcommand{\bea}{\begin{eqnarray}}
\newcommand{\eea}{\end{eqnarray}}
\newcommand{\be}{\begin{equation}}
\newcommand{\ee}{\end{equation}}
\newcommand{\ba}{\begin{align}}
\newcommand{\ea}{\end{align}}

\newcommand{\bigO}{\mathcal{O}}

\newcommand{\w}[1]{\mbox{$\W_\infty[#1]$}}

   \makeatletter
  \let\over=\@@over \let\overwithdelims=\@@overwithdelims
  \let\atop=\@@atop \let\atopwithdelims=\@@atopwithdelims
  \let\above=\@@above \let\abovewithdelims=\@@abovewithdelims
\renewcommand\section{\@startsection {section}{1}{\z@}%
                                   {-3.5ex \@plus -1ex \@minus -.2ex}
                                   {2.3ex \@plus.2ex}%
                                   {\normalfont\large\bfseries}}

\renewcommand\subsection{\@startsection{subsection}{2}{\z@}%
                                     {-3.25ex\@plus -1ex \@minus -.2ex}%
                                     {1.5ex \@plus .2ex}%
                                     {\normalfont\bfseries}}

\newcommand{\pd}[2]{\frac{\partial #1}{\partial #2}}

\newcommand{\beq}{\begin{equation}}
\newcommand{\eeq}{\end{equation}}
\newcommand{\beqa}{\begin{eqnarray}}
\newcommand{\eeqa}{\end{eqnarray}}
\newcommand{\beqar}{\begin{eqnarray*}}

\newcommand{\ve}{{\varepsilon}}

\def\[{\big[}
\def\]{\big]}
\def\la{\langle}
\def\ra{\rangle}

\def\e{{\epsilon}}
\def\ve{{\varepsilon}}

\def\g{{\gamma}}

\def\a{{\alpha}}
\def\b{{\beta}}

\def\w{\omega}
\def\bz{{\bar z}}

\def\be{{\bar \epsilon}}

\def\bw{{\bar w}}

\def\CJ{{\mathcal J}}

\def\CM{{\mathcal M}}

\def\CO{{\mathcal O}}

\def\CS{{\mathcal S}}

\newcommand{\bra}[1]{\langle\,   #1\,    |}
\newcommand{\ket}[1]{ |\,   #1 \,  \rangle}

\newcommand{\dt}{{\text d}}

\def\+{{(+)}}
\def\-{{(-)}}
\def\0{{(0)}}
\def\1{{(1)}}
\def\2{{(2)}}
\def\3{{(3)}}

\newcommand{\bd}[1]{\begin{fmffile}{#1}\begin{fmfgraph*}}
\newcommand{\ed}{\end{fmfgraph*}\end{fmffile}}

\begin{document}
\begin{titlepage}
\unitlength = 1mm
\ \\
\vskip 3cm
\begin{center}

{\LARGE{\textsc{Loop-Corrected Virasoro Symmetry of 4D Quantum Gravity }}}

\vspace{0.8cm}
Temple He, Daniel Kapec, Ana-Maria Raclariu and Andrew Strominger

\vspace{1cm}

{\it  Center for the Fundamental Laws of Nature, Harvard University,\\
Cambridge, MA 02138, USA}

\vspace{0.8cm}

\begin{abstract}
Recently a boundary energy-momentum tensor $T_{zz}$  has been constructed from the soft graviton operator for any 4D quantum theory of gravity in asymptotically flat space.  Up to an ``anomaly'' which is one-loop exact, $T_{zz}$ generates a Virasoro action on the 2D celestial sphere at null infinity. Here we show by explicit construction that the effects of the IR divergent part of the anomaly can be eliminated by a one-loop renormalization that shifts $T_{zz}$. 

 \end{abstract}

\vspace{1.0cm}

\end{center}

\end{titlepage}

\pagestyle{empty}
\pagestyle{plain}

\def\vx{{\vec x}}
\def\p{\partial}
\def\po{$\cal P_O$}

\pagenumbering{arabic}
 

\tableofcontents

\section{Introduction}

Recently it has been demonstrated \cite{Kapec:2014opa} that any theory of gravity in four asymptotically flat dimensions has, at tree-level, a Virasoro or ``superrotational'' symmetry that acts on the celestial sphere at null infinity.  This verified conjectures in \cite{deBoer:2003vf,Banks:2003vp,Barnich:2009se,Barnich:2010eb,Barnich:2011ct} and follows from the newly discovered subleading soft graviton theorem \cite{Cachazo:2014fwa}. Except for an anomaly arising from one-loop exact infrared (IR) divergences \cite{Bern:2014vva,Bern:2014oka,Broedel:2014bza,Bianchi:2014gla,He:2014bga}, this subleading soft theorem extends to the full quantum theory. However, the implications of this anomaly for the Virasoro symmetry of the full quantum theory are not understood, and the exploration of such implications comprises the subject of the present paper. 

There are several open possibilities. One is that the Virasoro action is defined in the classical but not in the quantum theory. If so, anomalous symmetries still have important quantum constraints that would be interesting to understand. A second possibility is that the Virasoro action acts on the full quantum theory, but that the generators and symmetry action are renormalized at one-loop. This is suggested by the discussion in \cite{hps}, where it is pointed out that the very definition of a scattering problem in asymptotically flat gravity requires an infinite number of exactly conserved charges and associated symmetries, as well as by \cite{Cachazo:2014dia}, which found that the anomaly vanishes with an alternate order of soft limits. A third possibility is that the implications can only be properly formulated in a Faddeev-Kulish \cite{ Chung:1965zza,Kibble:1969ip,Kibble:1969ep,Kibble:1969kd,Kulish:1970ut,Kulish:1971jb} basis of states (constructed for gravity in \cite{Ware:2013zja}), in which case all IR divergences are absent. After all, IR divergences preclude a Fock-basis $\CS$-matrix for quantum gravity and, although we have become accustomed to ignoring this point,  it is hard to discuss symmetries of an object which exists only formally! 

In this paper we give evidence which is consistent with, but does not prove, the second hypothesis, which states that the Virasoro action persists to the full quantum theory but requires the generators to have a one-loop correction.  We use the recent construction of a 2D energy-momentum tensor $T_{zz}$ found in \cite{Kapec:2016jld,Cheung:2016iub}, where $z$ is a coordinate on the celestial sphere, in terms of soft graviton modes. The tree-level subleading soft theorem \cite{Cachazo:2014fwa} implies that insertions of $T_{zz}$ in the tree-level $\CS$-matrix infinitesimally generate a Virasoro action on the celestial sphere.  At one-loop order, the subleading soft theorem has an IR divergent term with a known universal form. This spoils the Virasoro action generated by $T_{zz}$ insertions. However, we show explicitly that the effects of the IR divergent term can be removed by a certain shift in $T_{zz}$ that is quadratic in the soft graviton modes. The possibility that this could be achieved by a simple shift is far from obvious and requires a number of nontrivial cancellations. 

This does not demonstrate that there is a Virasoro action on the full quantum theory generated by a renormalized energy-momentum tensor, as there may also be an IR finite one-loop correction to  the subleading soft theorem. At present, little is known about such finite corrections. In all cases which have been analyzed \cite{He:2014bga,Bern:2014oka}, the finite part of the correction vanishes. Yet, there is no known argument that this should always be the case, and this remains an open issue for us.

The outline of this paper is as follows. In section two we fix conventions and recall the construction of the tree-level soft graviton energy-momentum tensor. We then reproduce the derivation of the one-loop exact IR divergent corrections to the subleading soft graviton theorem in section three.  Finally, this divergence is rewritten in section four in terms of the formal matrix element of another quadratic soft graviton operator, effectively renormalizing the tree-level energy-momentum tensor.

\section{Tree-Level Energy-Momentum Tensor}
In this section, we review the derivation of the 2D tree-level energy-momentum tensor living on the celestial sphere at  null infinity \cite{Kapec:2016jld,Cheung:2016iub}. 
Asymptotic one-particle states are denoted by $| p, s\ra$, where $p$ is the 4-momentum and $s$ is the helicity, and such states are normalized so that
\begin{align}
	\la p', s' |p,s\ra = (2\pi)^3 \big(2p^0 \big)\delta_{ss'}\delta^{(3)}(\vec p - \vec p')\ .
\end{align}
  An $n$-particle $\mathcal{S}$-matrix element is denoted by
\begin{align}
\begin{split}
	\CM_n &\equiv \la \text{out}|\CS|\text{in}\ra\ ,
\end{split}
\end{align}
where $|\text{in}\ra \equiv |p_1, s_1; \cdots; p_m,s_m\ra$ and $\la \text{out}| \equiv \la p_{m+1}, s_{m+1}; \cdots; p_n,s_n|$. Consider the amplitude 
\begin{equation}
\mathcal{M}^{\pm}_{n+1}(q)\equiv \la \text{out};q,\pm 2|\CS|\text{in}\ra\ ,
\end{equation}
consisting of $n$ external hard particles along with an additional external graviton that has momentum $p_{n+1} \equiv q$, energy $p_{n+1}^0\equiv\omega$, and polarization $\ve_{\mu \nu}^{\pm}$. Denoting the same amplitude without the extra external graviton as $\mathcal{M}_n$, the tree-level soft graviton theorem states that
\begin{equation}
\lim_{\omega\to 0}\mathcal{M}^{\pm}_{n+1}(q)= \left[	S_n^{(0)\pm} + S_n^{(1)\pm}	+\mathcal{O}(q)	\right]\mathcal{M}_n\ ,
\end{equation}
where the leading and subleading soft factors are given by
\begin{equation}
\begin{split}
S_n^{(0)\pm} = \frac{\kappa}{2} \sum_{k=1}^n \frac{ p_k^\mu p_k^\nu  \ve^\pm_{\mu\nu} (q)  }{ p_k \cdot q }\ , \qquad S_n^{(1)\pm} = - \frac{i\kappa}{2} \sum_{k=1}^n \frac{ \ve^{\pm}_{\mu\nu}  (q) p_k^\mu  q_\lambda  }{ p_k \cdot q } \CJ_k^{\lambda\nu}\ , 
\end{split}
\end{equation}
respectively. Here, $\kappa \equiv \sqrt{32\pi G}$ is the gravitational coupling constant, and $\CJ_k^{\lambda\nu}$ is the total angular momentum operator for the $k$th particle.

Asymptotically flat metrics in Bondi gauge take the form
\begin{equation}
\begin{split}
ds^2 &= - du^2 - 2 du dr + 2r^2\g_{z\bz} dz d\bz \\
&\qquad \qquad \qquad + \frac{2m_B}{r} du^2 + r C_{zz} dz^2 + r C_{\bz\bz} d\bz^2 + D^zC_{zz} du dz + D^{\bz}C_{\bz\bz} du d\bz + \cdots\ . 
\end{split}
\end{equation}
Here, $\gamma_{z\bz}\equiv \frac{2}{(1+z\bz)^2}$ is the round metric on the $S^2$ and $D_z$ is the associated covariant derivative. The coordinates $(u,r,z,\bz)$ are asymptotically related to the standard Cartesian coordinates according to 
\begin{equation}\label{cd}
\begin{split}
x^0 = u + r\ , \qquad x^i = r {\hat x}^i(z,\bz)\ , \qquad {\hat x}^i(z,\bz) = \frac{1}{1+z\bz} \left( z + \bz , - i ( z - \bz ) ,1 - z \bz \right)\ . 
\end{split}
\end{equation}
A massless particle with momentum $p_k$  crosses the celestial sphere at a point $(z_k,\bz_k)$. In the helicity basis, the particle momentum and polarization can be parameterized by an energy $\omega_k$ and this crossing point. It follows from \eqref{cd} that
\begin{equation}
\begin{split}\label{momexplicit}
p_k^\mu &= \omega_k \left( 1 , \frac{z_k + \bz_k}{1 + z_k \bz_k} , \frac{ - i ( z_k - \bz_k )}{1 + z_k \bz_k} , \frac{1 - z_k \bz_k}{1 + z_k \bz_k} \right)\ , \quad k=1,\cdots,n\ , \\
\ve^+_\mu( p_k ) &=  \frac{1}{\sqrt{2}} \left( - \bz_k , 1 , - i , - \bz_k \right)\ ,  \qquad \ve_\mu^-(p_k )  =    \frac{1}{\sqrt{2}} \left( - z_k , 1 ,   i , - z_k \right)\ ,\\  
q^\mu(z) &= \omega \left( 1  , \frac{z  + \bz}{1 + z  \bz}  , \frac{- i ( z  - \bz  )}{1 + z  \bz} , \frac{1 - z  \bz}{1 + z  \bz}   \right)\equiv \omega \hat{q}^\mu(z)\ , \\
\ve^+_\mu(q) &=  \frac{1}{\sqrt{2}} \left( - \bz , 1 , - i , - \bz \right)\ , \qquad \ve_\mu^-(q)  =    \frac{1}{\sqrt{2}} \left( - z , 1 ,   i , - z \right)\ ,
\end{split}
\end{equation}
where the soft graviton polarization tensor is taken to be $\varepsilon_{\mu\nu}^{\pm}=\varepsilon^{\pm}_\mu \varepsilon^{\pm}_\nu$. 

Now, the perturbative fluctuations of the gravitational field have a mode expansion given by
\begin{equation}
\begin{split}
h^{\text{out}}_{\mu\nu} \big( x^0,\vec{x}\, \big) = \sum_{\a=\pm} \int \frac{d^3q}{(2\pi)^3} \frac{1}{2\omega_q} \left[ {\bar \ve}^{\a}_{\mu\nu} (q) a^{\text{out}}_\a ( q ) e^{ i q \cdot x } + \ve_{\mu\nu}^{\a} (q) a^{\text{out}}_\a (q)^\dagger e^{- i q \cdot x } \right]\ , 
\end{split}
\end{equation}
where $a^{\text{out}}_\a(q)^\dagger$ and $a^{\text{out}}_\a(q)$ are the standard creation and annihilation operators for gravitons obeying the commutation relations
\begin{equation}
\begin{split}
\big[   a^{\text{out}}_\a ( p )   , a^{\text{out}}_{\b} (q )^\dagger \big] = \left( 2\pi \right)^3 (2\omega)  \delta_{\a\b} \delta^{(3)} \left( \vec{p} - \vec{q}\,\right)\ . 
\end{split}
\end{equation}
The transverse components of the metric fluctuations near \ip are given by
\begin{equation}
\begin{split}
	C_{\bz\bz}(u,z,\bz) \equiv \kappa \lim_{r\to\infty} \frac{1}{r} \p_{\bz} x^\mu \p_{\bz} x^\nu h_{\mu\nu}^{\text{out}} \big( u + r , r{ \hat x}(z,\bz) \big)\  .
\end{split}
\end{equation}
The large-$r$ saddle-point approximation yields
\begin{equation}
\begin{split}\label{Czzexp}
C_{\bz\bz} (u,z,\bz) &=- \frac{i\kappa}{8\pi^2} {\hat \ve}_{\bz\bz} \int_0^\infty d\omega_q \left[ a^{\text{out}}_{-} \big( \omega_q {\hat x}\big) e^{- i \omega_q u } - a^{\text{out}}_{+} ( \omega_q {\hat x} \big)^\dagger e^{i \omega_q u} \right]\ ,
\end{split}
\end{equation}
where
\begin{equation}
\begin{split}
{\hat \ve}_{\bz\bz} = \frac{1}{r^2} \p_{\bz} x^\mu \p_{\bz} x^\nu \ve_{\mu\nu}^+ (q )  = \frac{2}{(1+z\bz)^2}\ .
\end{split}
\end{equation}
Note that \eqref{Czzexp} is an intuitively plausible result since it states that the graviton field operator at a point $(z,\bz)$ on the celestial sphere has an expansion in plane wave modes whose momenta are aimed towards that point.

The Bondi news tensor $N_{zz}=\p_u C_{zz}$ has  Fourier components 
\begin{equation}
\begin{split}\label{Nzbzbomegadef}
	N_{zz}^\omega  \equiv \int du e^{i\omega u} N_{zz}\ , \qquad N_{\bz\bz}^\omega  \equiv \int du e^{i\omega u} N_{\bz\bz}\ . 
\end{split}
\end{equation}
The zero mode of the news tensor is defined by
\begin{equation}\begin{split}\label{tree_soft}
	N_{\bz\bz}^\0  &\equiv  \frac{1}{2} \lim_{\omega \to 0} \left( N_{\bz\bz}^\omega  + N_{\bz\bz}^{-\omega} \right) =  - \frac{\kappa}{8\pi} {\hat \ve}_{\bz\bz} \lim_{\omega \to 0} \left[ \omega a_{-}^{\text{out}}  \big( \omega {\hat x} \big) + \omega a_{+}^{\text{out}} \big( \omega {\hat x} \big)^\dagger   \right]\ .
\end{split}\end{equation}
Similarly, the first zero energy moment of the news is defined by
\begin{equation}\label{zeromodedef} 
\begin{split}
	N_{\bz\bz}^\1 &\equiv - \frac{i}{2} \lim_{\omega \to 0} \p_\omega \left( N_{\bz\bz}^\omega   - N_{\bz\bz}^{-\omega} \right) =  \frac{i\kappa}{8\pi}  {\hat \ve}_{\bz\bz}  \lim_{\omega \to 0} \big( 1 + \omega \p_\omega \big)   \left[   a_{-}^{\text{out}} \big( \omega {\hat x} \big)  -    a_{+}^{\text{out}} \big( \omega {\hat x} \big)^\dagger  \right]\ . 
\end{split}
\end{equation}
All of these quantities have nonvanishing $\CS$-matrix insertions even as $\omega\to 0$. The operator $N_{\bz\bz}^\0$ projects onto the leading Weinberg pole in the soft graviton theorem\cite{Weinberg:1965nx}, so its matrix elements are tree-level exact and are given by 
\begin{equation}\label{rde}
\begin{split}
 \bra{\text{out}} N_{\bz\bz}^\0 \CS \ket{\text{in}} &= -\frac{\kappa }{8\pi}\hat{\ve}_{\bz\bz}\lim_{\w\to 0} \w S^{\0-}_{n}\bra{\text{out}}  \CS \ket{\text{in}}\ ,
\end{split}
\end{equation}
where
\begin{align}\label{leadingS}
\begin{split}
	S^{\0-}_{n} = - \frac{\kappa}{2\omega} \left(1+z\bz \right) \sum_{k=1}^{n} \frac{\omega_k ( z - z_k ) }{ (\bz - \bz_k ) ( 1 + z_k \bz_k ) }\ .
\end{split}
\end{align}
In a similar fashion, $N_{\bz\bz}^\1$ projects onto the subleading $\bigO(1)$ term in the soft graviton theorem. At tree-level, its matrix elements are given by
\begin{align}\label{zeromodeins2}
\begin{split}
	\bra{\text{out}} N_{\bz\bz}^\1 \CS \ket{\text{in}} = \frac{i\kappa}{8\pi}\hat\ve_{\bz\bz}S^{\1-}_{n}\bra{\text{out}}  \CS \ket{\text{in}}\ ,
\end{split}
\end{align}
where
\begin{align}\label{subleadingS}
\begin{split}
	S^{\1-}_{n} = \frac{\kappa}{2}\sum_{k=1}^{n}  \frac{ (z - z_k  )^2 }{ \bz - \bz_k  } \left[ \frac{2 h_k }{z - z_k} -   \Gamma^{z_k}_{z_kz_k} h_k  - \p_{z_k}     +  |s_k| \Omega_{z_k}  \right].
\end{split}
\end{align}
In this expression, $\Gamma^{z}_{zz}$ is the connection on the asymptotic $S^2$, $h_k$ and $\bar{h}_k$ are the conformal weights given by
\begin{equation}
\begin{split}
	h_k \equiv \frac{1}{2} \left( s_k - \omega_k \p_{\omega_k} \right) ~, \qquad {\bar h}_k \equiv \frac{1}{2} \left( - s_k - \omega_k \p_{\omega_k} \right)\ ,
\end{split}
\end{equation}
and $\Omega_{z}$ is the corresponding spin connection.\footnote{The zweibein is $\big( e^+ , e^- \big)  = \sqrt{2\g_{z\bz}} \big( \dt z  , \dt \bz \big) $ and $\Omega^\pm{}_\pm = \pm \frac{1}{2} \big(\Gamma^z_{zz} \dt z - \Gamma^{\bz}_{\ \bz\bz} \dt\bz \big)$.}
As was demonstrated in \cite{Kapec:2016jld}, \eqref{zeromodeins2} implies that insertions of the operator
\begin{equation}
\begin{split}\label{Tzzdef}
	T_{zz} \equiv \frac{4i}{\kappa^{2}} \int d^2 w \frac{\gamma^{w\bw}}{z - w } D_w^3 N^\1_{\bw\bw}\  
\end{split}
\end{equation}
into the tree-level $\mathcal{S}$-matrix reproduce the Ward identity for a 2D conformal field theory:
\begin{equation}
\begin{split}\label{stresstensorins}
	& \la\text{out}| T_{zz}\CS|\text{in}\ra =  \sum_{k=1}^{n} \left[   \frac{  h_k }{ ( z - z_k )^2 } + \frac{ h_k }{z - z_k }\Gamma^{z_k}_{z_kz_k}    +  \frac{1}{z - z_k }   \left( \p_{z_k} -   |s_k| \Omega_{z_k}   \right)  \right]  \la\text{out}| \CS|\text{in}\ra\ .
\end{split}
\end{equation}

\section{One-Loop Correction to the Subleading Soft Graviton Theorem}
The matrix element \eqref{rde} is exact because the leading Weinberg pole in the soft graviton expansion is uncorrected. On the other hand, the subleading theorem which governs the $\CO(1)$ terms in the soft graviton expansion does have quantum corrections that modify the matrix element \eqref{zeromodeins2} \cite{Bern:2014oka}. These corrections are known to be one-loop exact, and arise from IR divergences in soft exchanges between external lines. Indeed, they must be present in order to cancel (within suitable inclusive cross-sections) IR divergences that arise from the Weinberg pole. The divergent part of this one-loop correction was derived in \cite{Bern:2014oka}, which we will we now review.

The loop expansion of the $n$-particle scattering amplitude is 
\begin{equation} \label{expansion} \mathcal M_n=
	\sum_{\ell=0}^\infty \mathcal M_n^{(\ell)} \kappa^{2\ell}\ ,
\end{equation}
where we factored out the $\kappa^2$ term that comes along with each additional loop.\footnote{In addition, there is a factor of $\kappa^{n-2}$ in each $\mathcal M_n^{(\ell)}$ due to the $n$ external lines.}
In dimensional regularization with $d=4-\e$, the divergent part of the one-loop $n$-point graviton scattering amplitudes is universally related to the tree amplitude according to \cite{Dunbar:1995ed,Naculich:2011ry}
\begin{align}\label{1-loop}
	\left.\mathcal M_n^{(1)} \right|_{\text{div}}= \frac{\sigma_n}{\e}\mathcal M_n^{(0)} \ ,
	\end{align}
with 
\begin{align}\label{sigma}
	\sigma_n \equiv -\frac{1}{4(4\pi)^2}\sum_{i,j=1}^n (p_i \cdot p_j) \log\frac{\mu^2}{-2p_i \cdot p_j} \ .
\end{align}
The $\mathcal{O}\big(\epsilon^{-1}\big)$ singularity is  due exclusively to IR divergences because pure gravity is on-shell one-loop finite in the UV and has no collinear divergences. 
Using \eqref{1-loop} and applying the tree-level soft theorem involving a negative-helicity soft graviton, we obtain
\begin{align}\label{soft_thm}
\begin{split}
	\left.\mathcal M_{n+1}^{(1)-}(q) \right|_{\text{div}}\xrightarrow{q \to 0}  \frac{\sigma_{n+1}}{\e} \left(S_{n}^{(0)-} + S_{n}^{(1)-} \right) \CM_{n}^{(0)} \ .
\end{split}
\end{align}
We would like to expand the above equation in powers of the soft energy $\omega$. To proceed, we separate $\sigma_{n+1}$ into two terms, one with the soft graviton momentum $ q$ and one without: \begin{align}\label{sigma'}
\begin{split}
	\sigma_{n+1} = \sigma_{n} + \sigma_{n+1}'\ , \quad \sigma_{n+1}' \equiv - \frac{1}{2(4\pi)^2} \sum_{i=1}^{n} (p_i \cdot q)\log\frac{\mu^2}{-2p_i \cdot q} \ .
\end{split}
\end{align} 
Note that $\sigma_{n+1}' = \bigO(\w)$ as the $\log \omega$ term vanishes by momentum conservation, while $\sigma_{n} = \bigO\big( \w^0 \big)$. We then find, up to $\bigO\big(\w^0\big)$,
\begin{equation}\label{uni}
	\left. \mathcal M_{n+1}^{(1)-} \right|_{\text{div}}\xrightarrow{q \to 0}  \left(S_{n}^{(0)-} + S_{n}^{(1)-} \right) \left.\CM_{n}^{(1)}\right|_{\text{div}}+\frac{\sigma'_{n+1}}{\e}S_{n}^{(0)-}\CM_{n}^{(0)} - \frac{1}{\e}\left(S_{n}^{(1)-} \sigma_{n}\right) \mathcal{M}_{n}^{(0)}  \ ,
\end{equation}
where $S_n^{(1)-}$ in the last term acts only on the scalar $\sigma_{n}$. The anomalous term consists of the last two terms on the right-hand-side of the above equation and is $\bigO\big( \w^0 \big)$. It is a universal correction to the subleading soft theorem from IR divergences.

Thus far, we have been focusing on the IR divergent part of the one-loop amplitude. However, the one-loop amplitude also has a finite piece:
\begin{equation}
\mathcal M_n^{(1)}= \left. \mathcal M_n^{(1)} \right|_{\text{div}}+ \left.\mathcal M_n^{(1)} \right|_{\text{fin}}\ .
\end{equation}
It is expected from \cite{Bern:2014oka} that
\begin{equation}\label{fin}
	\left.\mathcal M_{n+1}^{(1)-} \right|_{\text{fin}}\xrightarrow{q \to 0}  \left(S_{n}^{(0)-} + S_{n}^{(1)-} \right)\left. \CM_{n}^{(1)}\right|_{\text{fin}}+ \Delta_{\text{fin}} S_{n}^{(1)-} \CM_{n}^{(0)} \ ,
\end{equation}
where $\Delta_{\text{fin}} S_n^{(1)-}$ is the one-loop finite correction to the negative-helicity subleading soft factor $S_n^{\1-}$. Given that the subleading soft graviton theorem is one-loop exact \cite{Bern:2014oka}, it follows that the all-loop  soft graviton theorem  is 
\begin{align}\label{som}
\begin{split}
	 \mathcal M_{n+1}^{-}\xrightarrow{q \to 0}  \left[S_{n}^{(0)-} + S_{n}^{(1)-} + \kappa^2\left(\frac{\sigma'_{n+1}}{\e}S_{n}^{(0)-} - \frac{1}{\e}\left(S_{n}^{(1)-} \sigma_{n}\right) + \Delta_{\text{fin}} S_{n}^{(1)-} \right)\right] \CM_{n}\ ,
\end{split}
\end{align}
where the terms in square brackets proportional to $\kappa^2$ are the IR divergent and finite parts of the anomaly. 

Little appears to be currently known about $\Delta_{\text{fin}} S_{n}^{\1-}$. In all explicitly checked cases, including all identical helicity amplitudes and certain low-point single negative-helicity amplitudes, it was demonstrated that there are no IR finite corrections to the subleading soft graviton theorem \cite{He:2014bga,Bern:2014oka}, implying $\Delta_{\text{fin}} S_{n}^{(1)-}=0$ for these cases. Nevertheless, we are unaware of any argument indicating that this term always vanishes, or on the contrary that its form is universal.  In the absence of such information, we will restrict our consideration to the universal divergent correction  given in \eqref{uni}.

\section{One-Loop Correction to the Energy-Momentum Tensor}

The one-loop corrections \eqref{som} to the subleading soft factor are expected to result in corrections to the tree-level Virasoro-Ward identity \eqref{stresstensorins}. In this section, we show that this is indeed the case. Moreover, we find that the effects of the universal divergent correction in \eqref{uni} can be eliminated by a corresponding one-loop correction to the energy-momentum tensor. That is, whenever we have $\Delta_{\text{fin}} S_{n}^{(1)-}=0$, the shifted energy-momentum tensor obeys the unshifted Virasoro-Ward identity \eqref{stresstensorins}.

The tree-level matrix elements of the operator $N_{\bz\bz}^\1$ are given by  \eqref{zeromodeins2}. At one-loop level, the matrix elements acquire a divergent correction of the form
\begin{equation}
	\left.\la\text{out}| N_{\bw\bw}^{(1)}\CS|\text{in}\ra\right|_{\text{div}} =  \frac{i\kappa^3}{8\pi}\hat\ve_{\bz\bz} \lim_{\omega \to 0} \big( 1 + \omega \p_{\omega} \big) \left( \frac{\sigma'_{n+1}}{\e}S_n^{(0)-} - \frac{1}{\e}\left( S_n^{(1)-} \sigma_{n}\right) \right)\la\text{out}| \CS |\text{in}\ra\ .\end{equation}
It immediately follows from \eqref{Tzzdef} that the IR divergent one-loop correction to the $T_{zz}$ Ward identity is given by
\begin{align}\label{delta_T}
\begin{split}
	& \la\text{out}| \Delta T_{zz} \CS|\text{in} \ra \\
	& ~~~~~~~ = -\frac{\kappa}{2\pi\e}\int d^2w\,\frac{\gamma^{w\bw}}{z-w} D_w^3 \left[\hat\ve_{\bw\bw}\lim_{\w\to0}(1+\w\p_{\w})\left(  \sigma'_{n+1}S_n^{(0)-} -(S_n^{(1)-} \sigma_{n})\right) \right]\la\text{out}| \CS |\text{in}\ra \ ,
\end{split}
\end{align}
where
\begin{equation}\label{deltaTdef}
 \la\text{out}| T_{zz} \CS|\text{in} \ra|_{\text{div}}\equiv \la\text{out}| \Delta T_{zz} \CS|\text{in} \ra\ .
 \end{equation}
It is far from obvious, but nevertheless possible, to rewrite this in terms of the zero modes of the Bondi news. This computation is done explicitly in appendix \ref{app1}, and we find that $\Delta T_{zz}$ can be expressed as
\begin{equation}\label{final_Tzz}
 \Delta T_{zz}  =-  \frac{2}{\pi \kappa^2 \epsilon}\int d^2w\, \frac{\gamma^{w\bw}}{z-w} \left( 2N_{ww}^{(0)}D_wN_{\bw\bw}^{(0)} + D_w\left(N_{ww}^{(0)}N_{\bw\bw}^{(0)}\right) \right)\ .
\end{equation}
Hence, the shifted energy-momentum tensor, given by
\begin{equation}
	\tilde T_{zz} = T_{zz}- \Delta T_{zz}\ ,
\end{equation}
obeys the unshifted Ward identity 
\begin{equation}
\begin{split}\label{stresstensorins2}
	& \la\text{out}| \tilde T_{zz}\CS|\text{in}\ra =  \sum_{k=1}^n \left[   \frac{  h_k }{ ( z - z_k )^2 } + \frac{ h_k }{z - z_k }\Gamma^{z_k}_{z_kz_k}    +  \frac{1}{z - z_k }   \left( \p_{z_k} -   |s_k| \Omega_{z_k}   \right)  \right]  \la\text{out}| \CS|\text{in}\ra\  
\end{split}
\end{equation}
to all orders, whenever $\Delta_{\text{fin}} S_{n}^{\1-}=0$.

This result seems interesting for a number of reasons. First of all, note that while the renormalized soft factor contains logarithms and explicit dependence on the renormalization scale, such terms do not appear in the anomalous contribution to the energy-momentum tensor.  Furthermore, the fact that the divergence takes the form of a matrix element involving only the local operators  $N_{ww}^{(0)}(w)$ and $N_{\bw\bw}^{(0)}(w)$ allows us to perform an ``IR renormalization" of the operator $T_{zz}$ by subtracting away the divergent operator. The form of the divergence, when rewritten in terms of the soft graviton operators, is reminiscent of the forward limit of a scattering amplitude.  However, it remains to be seen whether or not there are finite corrections to the Ward identity \eqref{stresstensorins2} and, if so, whether or not they can be eliminated by a further finite shift of the energy-momentum tensor.

\section*{Acknowledgements}
We are grateful to D. Harlow, Z. Komargodski, P. Mitra,   B. Schwab, and A. Zhiboedov for useful conversations.   This work was supported in part by DOE grant DE-FG02-91ER40654.

\appendix

\section{IR Divergence of One-Loop $T_{zz}$ Correction}\label{app1}
In this appendix, we explicitly compute the matrix elements of $\Delta T_{zz}$ given in \eqref{delta_T} by
\begin{align}\label{delta_T-App}
\begin{split}
	& \la\text{out}| \Delta T_{zz} \CS|\text{in} \ra \\
	& ~~~~~~~ = -\frac{\kappa}{2\pi\e}\int d^2w\,\frac{\gamma^{w\bw}}{z-w} D_w^3 \left[\hat\ve_{\bw\bw}\lim_{\w\to0}(1+\w\p_{\w})\left(  \sigma'_{n+1}S_n^{(0)-} -(S_n^{(1)-} \sigma_{n})\right) \right]\la\text{out}| \CS |\text{in}\ra\ .
\end{split}
\end{align}
For completeness, we recall the expressions for the leading and subleading tree-level soft factors:
\begin{equation}
\begin{split}\label{softfactors}
	S_n^{(0)\pm} = \frac{\kappa}{2} \sum_{k=1}^n \frac{ p_k^\mu p_k^\nu  \ve^\pm_{\mu\nu} (q)  }{ p_k \cdot q }\ , \qquad S_n^{(1)\pm} = - \frac{i\kappa}{2} \sum_{k=1}^n \frac{ \ve^{\pm}_{\mu\nu}  (q) p_k^\mu  q_\lambda  }{ p_k \cdot q } \CJ_k^{\lambda\nu}\ . 
\end{split}
\end{equation}
Since $S^{(1)-}_n$ acts on a scalar in (\ref{delta_T-App}), the action of $\mathcal{J}_{k\mu\nu}$ is given by 
\begin{equation}
\begin{split}\label{angmomop}
\mathcal{J}_{k\mu\nu}\sigma_n &= - i \left[ p_{k\mu} \pd{}{p_k^\nu} - p_{k\nu} \pd{}{p_k^\mu } \right] \sigma_n\ .
\end{split}
\end{equation}
Using \eqref{softfactors}, \eqref{angmomop}, and momentum conservation, it follows that 
\begin{align}\label{soft-corr}
\begin{split}
\Delta_{\text{div}}S_n^{(1)-} &\equiv \frac{1}{\epsilon}\left[\sigma_{n+1}'S_n^{(0)-} - \left(S_n^{(1)-}\sigma_{n} \right) \right]  \\
&=\frac{\kappa}{4(4\pi)^2\epsilon} \sum_{i,j=1}^n\left[\frac{(p_i \cdot \varepsilon^- )^2}{ p_i\cdot q} (p_j\cdot q) \log\frac{ p_j\cdot q}{p_i\cdot p_j} - (p_i\cdot \varepsilon^-)(p_j\cdot \varepsilon^-)\log\frac{\mu^2}{-2p_i\cdot p_j} \right]\ .
\end{split}
\end{align}
Momentum conservation implies that \eqref{soft-corr} is independent of both the soft energy $\w$ and the renormalization scale $\mu$. It follows that \eqref{delta_T-App} becomes
\begin{equation}
\langle \text{out}|\Delta T_{zz} \mathcal{S}|\text{in}\rangle = -\frac{\kappa}{2\pi}\int d^2w \frac{\gamma^{w\bar{w}}}{z - w} D_{w}^3\left(\hat{\varepsilon}_{\bar{w}\bar{w}}\Delta_{\text{div}}S_n^{(1)-} \right)\la\text{out}| \CS |\text{in}\ra\ .
\end{equation}
Before proceeding, it is useful to define the quantity
\begin{equation}
\hat{\varepsilon}_{\bar{w}}\equiv\p_{\bw} \hat{x}^i(w)\varepsilon_{i}^+(q(w))=\frac{\sqrt{2}}{1+w\bw}\ ,
\end{equation}
so that $\hat{\varepsilon}_{\bar{w}\bar{w}}=\hat{\varepsilon}_{\bar{w}}\hat{\varepsilon}_{\bar{w}}$. It is then straightforward to show that
\begin{equation}\label{facts}
\begin{split}
	D_w^2\left(\hat{\varepsilon}_{\bar{w}}\varepsilon^- \right) &= 0\ , \\
	D_w^2 q &= 0\ , \\
	D^{\bw}D_w \left( \hat{\varepsilon}_{\bw\bw}\frac{(p_i\cdot \varepsilon^-)^2}{p_i\cdot \hat{q}}\right) &= -2\pi \omega_i \delta^{(2)}(w-z_i)\ ,
\end{split}
\end{equation}
where $ q^\mu=\omega  \hat q^\mu $. Using the first of the above identities, we have
\begin{equation}
D_{w}^3\left(\hat{\varepsilon}_{\bar{w}\bar{w}}(p_i\cdot \varepsilon^-)(p_j\cdot \varepsilon^-)\log\frac{\mu^2}{-2p_i\cdot p_j} \right) = 0\ ,
\end{equation}
which implies
\begin{equation}\label{correction1}
D^3_w\left(\hat{\ve}_{\bw\bw}\Delta_{\text{div}}S_n^{\1-}\right) = \frac{\kappa}{4(4\pi)^2\e}\sum^n_{i,j}D_w^3\left(\hat{\varepsilon}_{\bar{w}\bar{w}}\frac{(p_i \cdot \varepsilon^- )^2}{p_i\cdot q } (p_j\cdot q)  \log\frac{p_j \cdot q }{p_i\cdot p_j}\right)\ .
\end{equation}
To evaluate this, we distribute the covariant derivatives via the product rule and first compute the term
\begin{align}\label{term1}
\begin{split}
	&\sum_{i,j=1}^n D_w^3\left( \hat{\varepsilon}_{\bar{w}\bar{w}}\frac{(p_i \cdot \varepsilon^- )^2}{p_i\cdot q} (p_j\cdot q)\right)\log\frac{p_j\cdot q}{p_i\cdot p_j} = \\
	&~~~~~~~~~~~~ -2\pi\gamma_{w\bw}\sum_{i,j=1}^n\omega_i\left[(p_j \cdot \hat{q} )D_w\delta^{(2)}(w-z_i) + 3(p_j \cdot \partial_w \hat{q})\delta^{(2)}(w-z_i) \right]\log\frac{p_j \cdot \hat{q} }{p_j \cdot \hat{p}_i }\ ,
\end{split}
\end{align}
where we have used the last two identities of \eqref{facts} along with momentum conservation. Similarly, using \eqref{facts} and momentum conservation, we find
\begin{align}\label{Terms}
\begin{split}
	3\sum_{i,j=1}^n D_w\left[D_w\left(\hat{\varepsilon}_{\bar{w}\bar{w}}\frac{(p_i \cdot \varepsilon^- )^2}{p_i\cdot q} (p_j\cdot q) \right)D_w\left(\log\frac{ p_j\cdot q}{p_i\cdot p_j} \right) \right] &= 3\sum_{i,j=1}^nD_w \left[\hat{\varepsilon}_{\bar{w}\bar{w}}\frac{(p_i \cdot \varepsilon^- )^2}{p_i\cdot q }\frac{(p_j \cdot \partial_w q )^2}{ p_j\cdot q}\right]\ , \\
	\sum_{i,j=1}^n \hat{\varepsilon}_{\bar{w}\bar{w}}\frac{(p_i \cdot \varepsilon^- )^2}{p_i\cdot q } (p_j \cdot q ) D_{w}^3\left(\log\frac{p_j \cdot q }{p_i\cdot p_j}\right)  &=\sum_{i,j=1}^n \hat{\varepsilon}_{\bar{w}\bar{w}}\frac{(p_i \cdot \varepsilon^- )^2}{p_i \cdot q }  \frac{2}{(p_j\cdot q )^2}(p_j \cdot \partial_w q )^3\ .
\end{split}
\end{align}
Finally, using momentum conservation and the relationship between the soft momenta and polarization vectors \cite{Cheung:2016iub} 
\begin{equation}
 \varepsilon^+_\mu=\partial_w \left(\frac{1}{\sqrt{\gamma_{w\bar{w}}}}\hat{q}_{\mu} \right)\ ,
\end{equation}
we find
\begin{equation}\label{S+}
\sum_{j=1}^n \frac{\left(p_j \cdot \partial_w \hat{q} \right)^2}{p_j \cdot \hat{q} } = \sum_{j=1}^n \hat{\varepsilon}_{ww}\frac{(p_j \cdot \varepsilon^+ )^2}{p_j \cdot \hat{q} }\ .
\end{equation}
Substituting \eqref{term1} and \eqref{Terms} into \eqref{correction1}, and then using \eqref{rde} along with \eqref{S+}, we find
\begin{align}\label{TzzFin}
\begin{split}
\langle \text{out}| \Delta T_{zz}\mathcal{S}|\text{in}\rangle &= -\frac{2}{\pi\kappa^2\epsilon}\int d^2w\frac{\gamma^{w\bar{w}}}{z - w}\left\langle \text{out}\Big|\left[-2N_{\bar{w}\bar{w}}^{(0)} D_w N_{ww}^{(0)} + 3D_w\left(N_{ww}^{(0)}N_{\bar{w}\bar{w}}^{(0)} \right) \right]\mathcal{S} \ \Big|\text{in}\right\rangle \\
&= -\frac{2}{\pi\kappa^2\epsilon}\int d^2w\frac{\gamma^{w\bar{w}}}{z - w}\left\langle \text{out}\Big|\left[2N_{ww}^{(0)}D_w N_{\bar{w}\bar{w}}^{(0)}  + D_w\left(N_{ww}^{(0)}N_{\bar{w}\bar{w}}^{(0)} \right) \right]\mathcal{S} \ \Big|\text{in}\right\rangle\ ,
\end{split}
\end{align}
which is precisely \eqref{final_Tzz}.

\providecommand{\href}[2]{#2}\begingroup\raggedright

\end{document}